\documentclass[twocolumn,showpacs,preprintnumbers,amsmath,amssymb,prb]{revtex4}

\usepackage{amsmath}
\usepackage{amssymb}
\usepackage{graphicx}
\usepackage{epsfig}
\usepackage{dcolumn}
\usepackage{textcomp}
\usepackage{bm}
\usepackage[normalem]{ulem} 
\usepackage{color} 

\newcommand{\ket}[1]{|#1\rangle}
\newcommand{\bra}[1]{\langle #1|}

\begin{document}

\title{Spatial adiabatic passage in a realistic triple well structure} 

\author{J. H. Cole}
\affiliation{%
Centre for Quantum Computer Technology, School of Physics, The University of Melbourne, Melbourne, Victoria 3010, Australia.}
\affiliation{%
Institut f\"ur Theoretische Festk\"orperphysik
 and DFG-Center for Functional Nanostructures (CFN), Universit\"at Karlsruhe, 76128 Karlsruhe, Germany.}
\author{A. D. Greentree, L. C. L. Hollenberg}
\affiliation{%
Centre for Quantum Computer Technology, School of Physics, The University of Melbourne, Melbourne, Victoria 3010, Australia.}
\author{S. Das Sarma}
\affiliation{%
Condensed Matter Theory Center, Department of Physics, University of Maryland, College Park, Maryland 20742-4111, USA.}

\date{\today}
             
\begin{abstract}
We investigate the evolution of an electron undergoing Coherent Tunnelling via Adiabatic Passage (CTAP) using the solution of the one-dimensional Schr\"{o}dinger equation in both space and time for a triple-well potential.  We find the eigenspectrum and complete time evolution for a range of different pulsing schemes.  This also provides an example of a system which can be described with the tools from both quantum optics and condensed matter.  We find that while the quantum optics description of the process captures most of the key physics, there are important effects which can only be correctly described by a more complete representation.  This is an important point for applications such as quantum information processing or quantum control where it is common practice to use a reduced state space formulation of the quantum system in question. 
\end{abstract}

\maketitle

\section{Introduction}
The fabrication and control of mesoscopic quantum systems is of continued interest from both a fundamental and practical point of view.  One of the most exciting aspects of this is the convergence between quantum optics and quantum electronics.  While these have been traditionally treated as separate but related fields, the raft of new experimental and theoretical results have shown many important links between these topics.  This is especially true in quantum information processing and quantum control where the techniques and concepts used in one physical system are regularly applied to the other.  An important aspect of this convergence of quantum optics and electronics is the interoperability of the tools used to describe their respective systems.  

To explore the convergence between quantum optics and quantum electronics, we consider the recent analysis of STIRAP (Stimulated Raman Adiabatic Passage), which was originally developed in the quantum optics framework\cite{Vitanov:01}, applied to spatial coherence of electrons, in which case it is termed CTAP (Coherent Tunnelling via Adiabatic Passage)\cite{Greentree:04}.  This is an example of a non-trivial time dependent problem which has both a quantum optics and condensed matter version and which is of interest in both communities.  The CTAP protocol has been suggested as a means to transport electrons from one potential well to another via adiabatic manipulation of the ground state wavefunction.  The transport of spins using this protocol has been proposed as the basis of quantum computing schemes\cite{Hollenberg:06, Greentree:06}.  Similarly, transporting single atoms\cite{Eckert:04,Eckert:06} and Bose-Einstein condensates\cite{Graefe:06,Rab:07} within harmonic traps have also been investigated.  CTAP has recently been demonstrated using photons in triple-core optical waveguides, using modulated separations between the cores\cite{Longhi:06, Longhi:07,Longhi:07b}.

The development of the CTAP protocol was carried out directly from the quantum optics framework which translates to an ideal localization assumption. In this work we relax this assumption and investigate the protocol in a manner more appropriate to a realistic quantum nanoelectronic/solid-state setting, and highlight the differences and commonalities of the approaches to each physical context. The results of our analysis are directly applicable to the implementation in the quantum dot setting, including gate defined~\cite{Loss:98, vanderWiel:03,Gaudreau:06,Schroer:07,Rogge:08,GroveRasmussen:08,Simmons:07} and donor based Coloumb confined \cite{Greentree:04} single electron systems.

To this end, this paper necessarily focuses on describing few state quantum systems using tools common to either the quantum optics or condensed matter communities. The theoretical treatment of CTAP and other related schemes\cite{Fabian:05,Michaelis:06,Petrosyan:06,Emary:07,Kamleitner:08,Brandes:01,Siewert:06} that involve the coherent behaviour of electrons often employs the finite state approximation (FS), also variously known as the tight-binding or modal approximation.  As such, the electron is located at one of a number of localised sites and can tunnel coherently between these sites.  The evolution of the system is then solved using matrix mechanics using the finite state space.  In this model, the mathematics of the system evolution is often equivalent to a few state quantum optics problem, especially when considering donor electrons in semiconductors or quantum dots containing few electrons.  This approach has become particularly important given recent work on quantum control and quantum computing in both quantum optics and solid-state\cite{Nielsen:00,Chakrabarti:07}.

We investigate the generic problem of the link between the finite state model parameters with those that can be determined from conventional solid-state physics and the solution to the Schr\"{o}dinger equation in one dimension using wave mechanics.  We use CTAP as a specific example where the appropriate time variation of the potentials is calculated for a series of square wells separated by finite barriers.  We discuss the analysis of the resulting eigenspectrum and then perform 1D time dependent calculations to investigate the system evolution.  

At this point, it should be noted that the solution of the spatially varying wavefunction for an arbitrary 3D potential is the more sophisticated and ultimately correct treatment.  At best, the FS treatment is a physically motivated approximation for treating the shuttling of electrons between quantum dots or donor sites.  Here we choose a mid point between FS and full 3D, where the dynamics are dominated by the potential variation in only one of the three dimensions, the other two dimensions consisting of tight confining potentials.  Most of the essential physics involving barrier penetration and finite confinement enter the problem at this 1D level.  Further effects introduced in the full 3D problem are mostly system specific, depending on both the characteristics of the host lattice and the nature of the confining potentials.  Investigating the similarities and differences between the Schr\"{o}dinger wave (SW) description, even in 1D, and the finite state approximation (FS) provides important insight into the validity of this approximation.

While this paper focuses on solutions to the one dimensional Schr\"{o}dinger equation, the analysis and methodology applies to many real systems, as coherent tunnelling can often be treated as a one dimensional problem.  In GaAs quantum dots for instance, the separability of the spatial degrees of the wavefunction often allows the dynamics in one dimension to be treated independently of the confinement of the wavefunction in the other two dimensions~\cite{vanderWiel:03}.  A notable exception is the charge states associated with isolated dopant atoms in semiconductors, where the Coulombic nature of the trapping potential introduces divergences which must be handled with some care.  In this case, our analysis still provides some insight as similar qualitative behaviour is seen in these systems when compared to other forms of trapping potentials.

In the following section we discuss the links between the approaches used in quantum optics (the FS model) and those commonly used in condensed matter physics (the SW model).  We then introduce CTAP in section~\ref{sec:ctap} as an example of a nontrivial adiabatic process which can be modelled using either formalism.  As CTAP is an adiabatic process, much insight can be gained by plotting the eigenspectrum.  In section~\ref{sec:eigspec} we find the eigenenergies and wavefunctions using the SW formalism and use this to estimate the adiabatic time scale.  Using the FS description of CTAP, the population of the central well is predicted to be exactly zero in the adiabatic limit.  Using the SW formalism, section~\ref{sec:well2}, we can put quantitative bounds on this population.  Finally in section~\ref{sec:timeevol}, full time dependent evolution is calculated for the CTAP process and this is used to investigate the fidelity of transfer as a function of switching speed.  This allows direct comparison between the FS and SW descriptions.

\section{The Double well system and the link to quantum optics}\label{sec:doublewell}
The canonical system for investigating two-state quantum mechanics in the solid-state is the double well potential, whereas in atom optics it is the two-level atom in the rotating wave approximation.  Much of the mathematical properties of these two descriptions are trivially equivalent, which leads to them being often used almost interchangeably.  In this paper, we are primarily concerned with the points where this equivalence becomes non-trivial and ultimately breaks down, especially in the case of multi-well systems.  To motivate the discussion and introduce the relevant formalism, we will first point out several important features of the two-state and/or double well system as background.  

\begin{figure} [tb!]
\centering{\includegraphics[width=4cm]{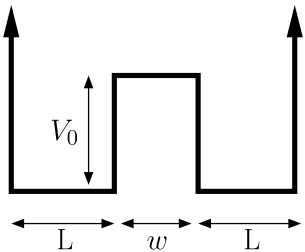}} 
\caption{The canonical two square well, finite barrier problem.\label{fig:diag_2well}}
\end{figure}

Consider a one dimensional system consisting of two potential wells separated by a finite potential barrier,~Fig.\ref{fig:diag_2well}.  We assume that the potential barrier between the wells is sufficiently high that each well contains at least one `bound'  state.  We also assume that the wells are narrow enough that the higher lying eigenstates of the system are sufficiently removed from the lowest two states that we can approximate the system as a two-state system.  The Hamiltonian governing the evolution of the lowest two eigenstates can be written as
\begin{equation}
H=\left(\begin{array}{cc}E_L & -\kappa \\ -\kappa & E_R \end{array} \right) 
\end{equation}
in a basis ($\ket{L}$ and $\ket{R}$) comprised of the ground states of the left- and right-hand wells respectively.  The coupling between states $\kappa$ is just a function of the height and width of the finite barrier and $E_L$ and $E_R$ correspond to the energies of each well defined separately, given an infinite barrier.  It is at this point that the analogy with a two-level atom in the rotating frame from quantum optics is made.  In a driven two level atom problem, the off-diagonal terms are given by the intensity of the driving field (the Rabi frequency) while the diagonal terms come from the energy mismatch between the bare transition and the frequency of the driving field (the detuning).  In the square well problem, the link between this reduced state space and the original spatially dependent wavefunction is then contained in the calculation of $\kappa$ for a finite barrier.

\begin{figure} [tb!]
\centering{\includegraphics[width=9cm]{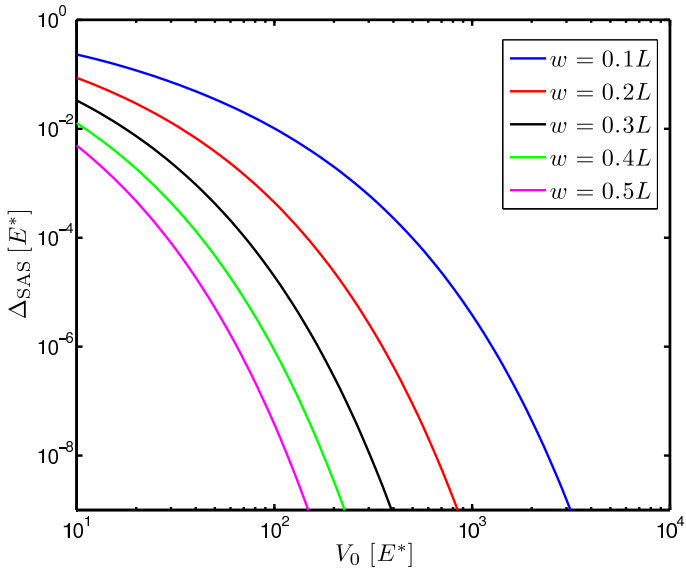}} 
\caption{The energy gap between the ground and excited states ($\Delta_{\rm{SAS}}$) as a function of the barrier height $V_0$ in units of $E^*$, for a double well system comprised of wells of width $L$ separated by a barrier of width $w$, as shown in Fig.~\ref{fig:diag_2well}.\label{fig:DeltaSAS}}
\end{figure}

The solution of the double well problem is most often constructed by matching boundary conditions of the solution to the Schr\"{o}dinger equation in each region separately\cite{Cohen:77}.  In the case of the double well, the solution does not have a closed form but can be evaluated numerically.  The calculation of $\kappa$ then reduces to finding the energy gap ($\Delta_{\rm{SAS}}$) between the two lowest (symmetric and antisymmetric) eigenstates of the double well system.  Fig.~\ref{fig:DeltaSAS} gives the value of $\Delta_{\rm{SAS}}$ for two wells of width $L$ separated by a barrier of width $w$ and height $V_0$.  The energy is expressed in normalized units ($E^*$) which corresponds to the ground state energy of an infinite well of width $L$.

A second method of solving the 1D Schr\"{o}dinger involves direct numerical discretization of the wavefunction, and will be used later for solving an arbitrary 1D potential problem.  If we discretize the region of interest into $n$ points $x_1,x_2,\dots, x_n$ separated by $\Delta x$, the wavefunction can then be computed at each point $\psi_k=\psi(x_k)$.  We then express the second derivative as a finite difference equation
\begin{equation}
\frac{\partial^2\psi(x_k)}{\partial x} \approx \frac{\psi_{k+1}-2\psi_k+\psi_{k-1}}{(\Delta x)^2}
\end{equation}
which allows us to construct the Hamiltonian of the system on a discrete lattice, 
\begin{equation}\label{eq:Hsw}
H_{\rm{SW}}(x_k)=V(x_k) - \frac{\psi_{k+1}-2\psi_k+\psi_{k-1}}{(\Delta x)^2}
\end{equation}
and express the Schr\"{o}dinger equation as a matrix equation.  For our purposes we truncate the matrix at either end, which corresponds to infinite potential boundary conditions, $V(x_0)=V(x_{n+1})=\infty$.  Diagonalizing this Hamiltonian matrix gives the eigenvalues, the lowest ones of which are good approximations to the system eigenenergies.  Similarly, the wavefunctions will be the eigenvectors corresponding to these eigenvalues.  While this method is not particularly sophisticated, it does allow the inclusion of arbitrary potential shapes which will be important later.  In addition, because the resulting matrix is tridiagonal, this diagonalization process can be performed quickly and efficiently with modern sparse solving routines\cite{Press:96}.  

\section{CTAP in a realistic potential}\label{sec:ctap}
To illustrate the link between the finite state and Schr\"{o}dinger wave approaches, we take as an example a recently proposed technique for moving electrons spatially using coherent, adiabatic evolution.  As this scheme is the direct electrical analogue of STIRAP, the existing analyses specifically uses the FS model.  Here we investigate the state evolution and pulse design for this system using the SW approximation.  The full treatment of the spatial extent of the electron is ultimately the more correct description as it does not assume that the electron is tightly bound to one of the sites included in the finite state basis set.

We begin by reviewing the FS treatment of CTAP.  Given a single electron which can be localised in one of three distinct spatial regions, we describe this in a basis of three states which are nominally the ground state of the electron at each of the three sites.  These spatial locations are analogous to energy levels in a multi-level atom in quantum optics.  Obviously the first approximation here is that the excited states of the quantum wells can be ignored in this process.  This means any physics derived from the population of these higher levels is essentially lost, but we will discuss this later.  

\begin{figure} [tb!]
\centering{\includegraphics[width=8cm]{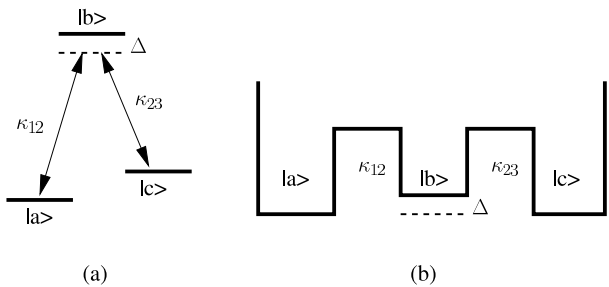}} 
\caption{Correspondence of the (a) quantum optics description of a three level atom with laser transitions which is equivalent to the finite state (FS) approximation and (b) a single electron undergoing coherent tunnelling between one of three finite wells (SW).  The tunnel matrix elements between states are $\kappa_{12}$ and $\kappa_{23}$, while the single-photon detuning (central well offset energy) is given by $\Delta$.\label{fig:QO_well_diag}}
\end{figure}

We assume that the tunnel barrier between the dots can be modulated in such a way that we may vary the coherent tunneling rates (or matrix elements) between dots.  This process is directly comparable to laser mediated transitions between levels (in the rotating frame) of a multi-level atom (see Fig.~\ref{fig:QO_well_diag}).  In the finite state model, the Hamiltonian is given by
\begin{equation}
H_{\rm{FS}} = \left(\begin{array}{ccc}0 & -\kappa_{12} & 0 \\ -\kappa_{12} & \Delta & -\kappa_{23} \\ 0 & -\kappa_{23} & 0 \end{array} \right) 
\end{equation}
where $\Delta$ is the offset energy (detuning) of the central site relative to the end sites, $\kappa_{12}$ and $\kappa_{23}$ are tunnel matrix elements and we have set the energies of the first and last sites equal.   

The eigenstates of this Hamiltonian are~\cite{Greentree:04,Vitanov:01}
\begin{eqnarray}
|{\cal D}_+\rangle &=& \sin \Theta_1 \sin \Theta_2 |a\rangle +
        \cos \Theta_2 |b\rangle +
        \cos \Theta_1 \sin \Theta_2 |c\rangle, \nonumber \\
|{\cal D}_-\rangle &=& \sin \Theta_1 \cos\Theta_2 |a\rangle -
        \sin \Theta_2 |b\rangle +
        \cos \Theta_1 \cos \Theta_2|c\rangle, \nonumber \\
|{\cal D}_0\rangle &=& \cos \Theta_1|a\rangle +
         0 |b\rangle -
        \sin \Theta_1 |c\rangle,
\label{eq:DressedStates}
\end{eqnarray}
where 
\begin{eqnarray}
\Theta_1 &=& \arctan \left(\kappa_{12}/\kappa_{23} \right), \nonumber
\\
\Theta_2 &=& \frac{1}{2} \arctan
\left[\left(\sqrt{(2\kappa_{12})^2 + (2\kappa_{23})^2}\right) / \Delta
\right].
\end{eqnarray}

The eigenenergies of these states are
\begin{eqnarray}
{\cal E}_\pm &=& \frac{\Delta}{2} \pm
\frac{1}{2}\sqrt{(2\kappa_{12})^2
+ (2\kappa_{23})^2 + \Delta^2}, \nonumber \\
{\cal E}_0 &=& 0. \label{eq:Eigenenergies}
\end{eqnarray}

Applying a specific pulse sequence to modulate the coherent tunnelling allows the transfer of the electron between the two extrema wells.  If the barrier closest to the electron is lowered first, the system adiabatically follows the ground state eigenstate spending some time in the central well.  This we refer to as the intuitive direction (ID).  Reversing the pulse sequence is termed the counter-inuitive direction (CID) and similarly results in a transfer of the electron but this time via the 1st excited state of the system.  This can be seen directly from the eigenstates Eq.~(\ref{eq:DressedStates}), where the CID path transforms from $\ket{a}$ at $t=0$ to $\ket{c}$ at $t=t_{\rm{max}}$ and hence the goal of the CID sequence is to maintain the system in $|{\cal D}_0\rangle$ throughout the evolution.

The original work on CTAP\cite{Greentree:04} assumed that the tunnelling rates between wells could be arbitrarily controlled, whereas here we are specifically concerned with how to modulate the \emph{barrier heights} to achieve an arbitrary tunnelling rate variation.  It is by investigating this link between the control of the barrier height and the tunnelling rates and associated issues that we can investigate the link between the FS and SW descriptions.  In this case, the SW version of the problem is described by a 1D potential consisting of three wells, separated by finite barriers (i.e.\ Fig.~\ref{fig:QO_well_diag}b).  The discrete form of the Hamiltonian of the system is then given by Eq.~(\ref{eq:Hsw}).  

If we start with the conventional CTAP pulsing scheme, the coupling between states is varied in time in a Gaussian fashion,
\begin{eqnarray}\label{eq:gaussianpulses}
\kappa_{12}(t) & = & \kappa_{\rm{max}}\rm{exp}\left\{ -\frac{[t-(t_{\rm{max}}+\sigma)/2]^2}{2\sigma^2} \right\} \\
\kappa_{23}(t) & = & \kappa_{\rm{max}}\rm{exp}\left\{ -\frac{[t-(t_{\rm{max}}-\sigma)/2]^2}{2\sigma^2} \right\},
\end{eqnarray}
where $\kappa_{\rm{max}}$ is the maximum coupling, $t_{\rm{max}}$ is the pulse time and $\sigma$ controls the width of the pulses in time.  To achieve an equivalent modulation using finite wells, we need to vary the barrier heights starting from some maximum barrier height ($V_{\rm{max}}$), down to some minimum height ($V_{\rm{min}}$) and back up to the maximum.  The CTAP pulses are then constructed by applying identical pulses to both barriers with some time delay, $\sigma$, between them.  

\begin{figure} [tb!]
\centering{\includegraphics[width=9cm]{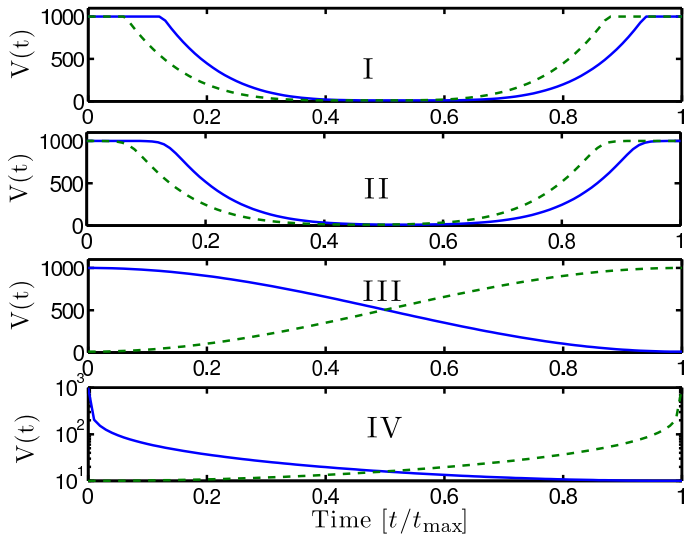}} 
\caption{Four different pulsing schemes for modulation of a barrier height to result in a CTAP transfer between wells.  In each case, the solid (dashed) curve is the height of barrier 1 (2) as a function of pulse time.  Note the log scale for pulse~\texttt{IV}.\label{fig:pulses}}
\end{figure}

To find an approximate barrier modulation pulse to replicate the behaviour of the Gaussian pulses, Eq.~(\ref{eq:gaussianpulses}), we can consider the conventional result that the probability of tunnelling through a finite (but large) barrier of height $V_0$ and width $w$ is proportional to $\mathrm{exp} [-w\sqrt{V_0}]$.  If we naively equate this to the coupling energy we obtain
\begin{equation}
\mathrm{exp} [-w\sqrt{V_0}] \propto \rm{exp}\left\{ -\frac{[t-(t_{\rm{max}}\pm\sigma)/2]^2}{2\sigma^2} \right\}
\end{equation}
and normalizing for the minimum and maximum barrier heights gives
\begin{equation}\label{eq:pulseI}
V(t)=\frac{V_{\rm{max}}-V_{\rm{min}}}{(t_{\rm{max}}+\sigma)^4}\left(2t - t_{\rm{max}} \mp \sigma \right)^4+V_{\rm{min}},
\end{equation}
which we designate as pulsing scheme $\texttt{I}$.  Fig.~\ref{fig:pulses}(a) shows the barrier heights given by Eq.~(\ref{eq:pulseI}) as a fraction of the pulse time for $\sigma=t_{\rm{max}}/16$, $V_{\rm{min}}=10$ and $V_{\rm{max}}=1000$. 

To derive a more rigorous pulsing scheme (\texttt{II}), we can take the previous calculation of coupling as a function of barrier height (Fig.~\ref{fig:DeltaSAS}) and use interpolation to calculate the required barrier height for a given coupling $\kappa$.  Fig.~\ref{fig:pulses}(b) shows this pulsing scheme for a barrier width of $w=0.2L$.  The functional form of this pulse is very similar to that of pulse \texttt{I} but with a smooth roll-off at the pulse maximum.  While pulse \texttt{II} is more difficult to calculate (as it requires the numerical solution to the finite well problem) this calculation need only be performed once with the pulses being then constructed via lookup tables and interpolation.

The pulses considered so far start and finish with the barriers both `high'.  It is also possible to construct a pulse where only one barrier is initially high, the other barrier (which is not required to confine the electron \emph{initially}) can be started in its `low' state\cite{Devitt:07}.  An example of this is
\begin{eqnarray}\label{eq:cospulses}
V_{1}(t) & = & (V_{\rm{max}}-V_{\rm{min}})\cos^2\left(\frac{\pi}{2} \frac{t}{t_{\rm{max}}}\right)+V_{\rm{min}} \\
V_{2}(t) & = & (V_{\rm{max}}-V_{\rm{min}})\sin^2\left(\frac{\pi}{2} \frac{t}{t_{\rm{max}}}\right)+V_{\rm{min}}, 
\end{eqnarray}
which we designate pulse \texttt{III}, see Fig.~\ref{fig:pulses}(c).  In this pulse scheme there no longer exists an ID pulse and the CID pulse corresponds to following the 2nd excited state of the system.  In a system in which the native barrier position is low and voltages need to be applied to raise the barrier height, an alternating pulse of this type may be simpler to implement.  It also does not require timing delays between the pulses, as the pulse overlap stems from the time reversal symmetry of the pulses.

While pulse \texttt{III} is relatively simple to describe and/or implement, it is not optimal in the sense that the greatest variation in barrier height is exactly where there is greatest overlap, resulting in a very narrow energy gap for adiabatic evolution.  A better pulsing scheme would be to construct pulses which display a trigonometric variation in \emph{coupling} rather than barrier height.  Again using Fig.~\ref{fig:DeltaSAS} to obtain such a pulse (\texttt{IV}) we obtain the pulse scheme shown in Fig.~\ref{fig:pulses}(d), which requires much finer control  in the initial and final stages of the pulse but results in a much smoother variation in the coupling over time and large adiabatic gap.
 
\section{Eigenspectra and Adiabatic Criteria}\label{sec:eigspec}
Using the method of section~\ref{sec:doublewell}, we can obtain the eigenenergies and eigenstates of the triple well system as a function of pulse time.  In this way we can construct the eigenspectrum, which provides information about the adiabaticity of the system, as well as determining the population distribution at each point in time.  While in principle the FS formalism provides this same information, using the SW method allows the inclusion of realistic spatially varying potentials.  This, in turn, allows calculation of the higher lying states and the spatial distribution of the wavefunction itself.

\begin{figure} [tb!]
\centering{\includegraphics[width=9cm]{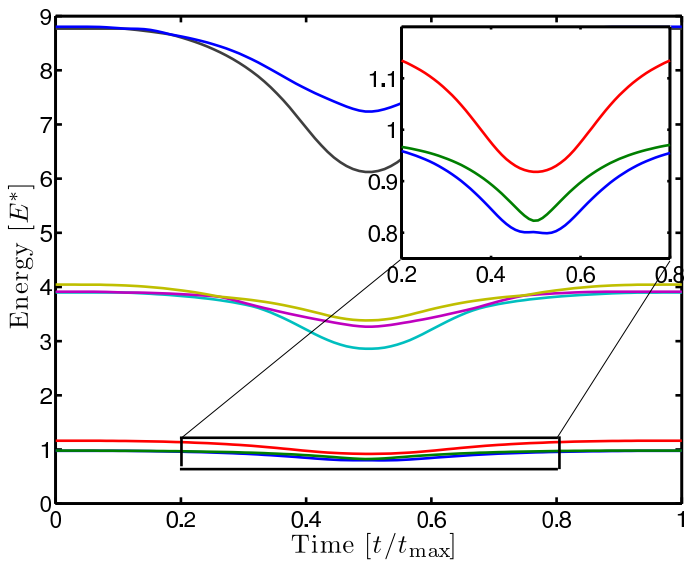}} 
\caption{Eigenspectrum of the three well system as a function of time within pulse sequence \texttt{II}.  The insert shows the three lowest eigenvalues in the central region of interest where the characteristics of the eigenstates are changing fastest.\label{fig:eigenspec1}}
\end{figure}

In Fig.~\ref{fig:eigenspec1} we show the eigenspectrum of the finite well system (consisting of 3 wells of width $L$ separated by two barriers of width $w=0.2L$) undergoing evolution due to modulation of the barrier heights according to pulse \texttt{II}.  In this plot, the manifolds of three states are visible at progressively higher energies.  Each of these manifolds corresponds to a bound state in a well of width $L$.  The width of the wells has been chosen so that there is a large separation between each group of three states which guarantees that the adiabatic evolution will leave the particle in the ground state of the occupied well at the end of the pulse sequence.

For this and subsequent calculations involving pulses \texttt{I} \& \texttt{II}, the bottom of the central potential has been raised such that $V(-L/2\leq x \leq L/2)=0.1E^*$, to separate the energies of the states within each manifold.  The sign of this offset does not qualitatively change the following results, provided that the energy of the lowest 3 eigenstates does not approach those of the next highest manifold of states.

For pulses \texttt{III} \& \texttt{IV}, the central well offset is unnecessary as the initial condition (with only one barrier high) naturally splits the degeneracy of the first three eigenstates.  In contrast, for all pulses sequences the left and right wells are identical, resulting in no offset between their ground state energies.  This has been shown to be an important requirement~\cite{Vitanov:01,Greentree:04,Kamleitner:08} as variation in energy between the initial and final well results in crossings in the eigenspectrum, which limits the adiabatic evolution.

\begin{figure} [tb!]
\centering{\includegraphics[width=9cm]{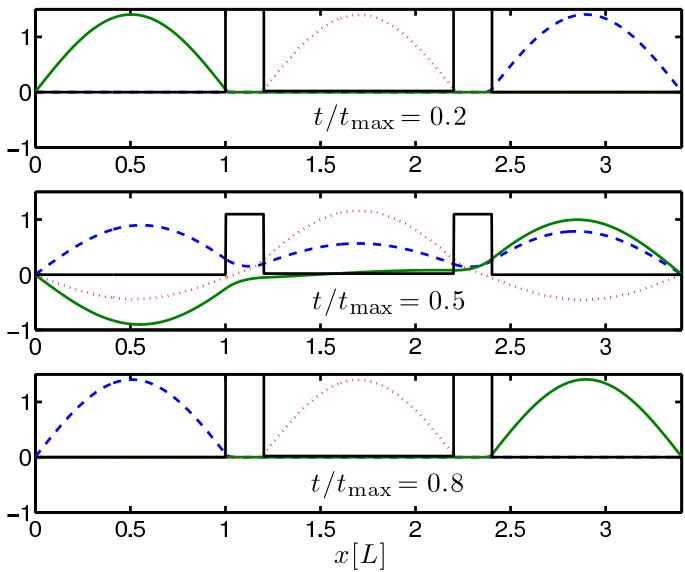}} 
\caption{Eigenstates of the three well system as a function of time within pulse \texttt{I} plotted at (a) $t/t_{\rm{max}}=0.2$, (b) $t/t_{\rm{max}}=0.5$ and (c) $t/t_{\rm{max}}=0.8$.  The eigenstates are the ground state (dashed), 1st excited (solid) and 2nd excited state (dotted) respectively, where the CTAP pulse corresponds to the 1st excited state.  Note, the potential is shown in arbitrary units for comparison.\label{fig:eigenstates}}
\end{figure}

Fig.~\ref{fig:eigenstates} shows the eigenstates of the three well system at three different points during pulse sequence \texttt{II}, whose eigenspectrum is shown in Fig.~\ref{fig:eigenspec1}.  The eigenstate corresponding to the CID pathway for CTAP is given by the 1st excited state (the solid line) in each subfigure.  Note that the electron is well localised at the beginning and end of the pulse, but the spatial location of the eigenstates have swapped.  At the halfway point of the pulse sequence, the eigenstates are superpositions of all three well states as well as having some non-negligible population within the barriers.  In contrast to the FS model the population in well 2 is not zero, even in the limits of long time and ideal coupling, due to the extended nature of the wavefunction.  However in common with the FS model, the central well population is strongly suppressed for the CID pathway state.  We discuss this in more detail in section~\ref{sec:well2}.

\begin{figure} [tb!]
\centering{\includegraphics[width=9cm]{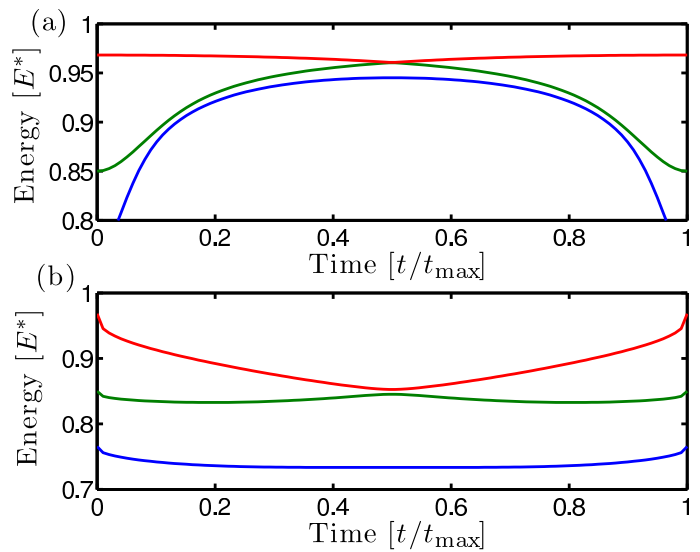}} 
\caption{Eigenspectrum of the three well system as a function of time for (a) pulse \texttt{III} and (b) pulse \texttt{IV}.  Note the very small energy gap for pulse \texttt{III}, due to the smaller tunnelling rate at the midpoint of the pulse sequence.\label{fig:pulses3and4}}
\end{figure}

We can perform the same analysis to obtain the eigenspectrum of the other pulse schemes, which are shown in Fig.~\ref{fig:pulses3and4}.  Note that, given the similarity between pulse \texttt{I} and \texttt{II}, the eigenspectrum from Fig.~\ref{fig:eigenspec1} is not presented here.  From Fig.~\ref{fig:pulses3and4} we immediately see that pulse \texttt{III} is indeed a poor choice as it results in a very small (but non-zero) gap between the first and second excited states.  In contrast, pulse \texttt{IV} provides a smooth variation in the energies of the system with a similar gap size to that resulting from pulses \texttt{I} and \texttt{II}.  

\begin{figure} [tb!]
\centering{\includegraphics[width=9cm]{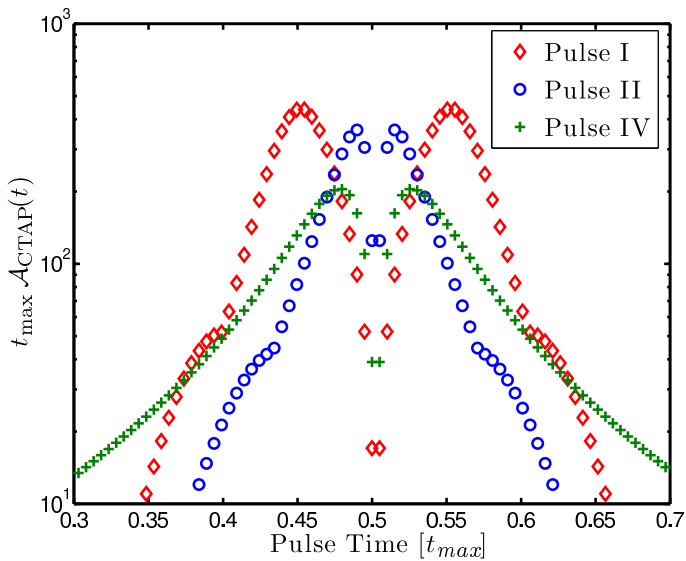}} 
\caption{$\mathcal{A}_{\rm{CTAP}}(t) $ plotted for three different pulse sequences (pulse \texttt{III} is not included as it requires excessively long pulse times).  From this plot we can estimate that for the chosen values of $V_{\rm{max}}$ and $V_{\rm{min}}$, the pulse time must be at least $t_{\rm{max}} \ge 1/100 (E^*)^{-1}$ to ensure adiabatic evolution.\label{fig:adiab}}
\end{figure}

Once we can calculate the eigenspectrum of the system, we can also estimate a time scale over which the system evolution is adiabatic.  To estimate this, we use
\begin{equation}
\mathcal{A}[\phi_i(t),\phi_j(t)]=\frac{\bra{\phi_j(t)}\frac{\partial H}{\partial t}\ket{\phi_i(t)}}{|\bra{\phi_j(t)}H\ket{\phi_j(t)}-\bra{\phi_i(t)}H\ket{\phi_i(t)}|^2}
\end{equation}
as the adiabaticity of a transfer between eigenstates $\phi_i$ and $\phi_j$\cite{Messiah:65}.  We can then define the adiabaticity of the CTAP process as
\begin{equation}
\mathcal{A}_{\rm{CTAP}}(t)=\rm{max}\{\mathcal{A}[\phi_1(t),\phi_2(t)],\mathcal{A}[\phi_2(t),\phi_3(t)]\}
\end{equation}
where the $\phi$'s are the first three eigenstates of the system (ordered by energy) at a given time $t$ in the pulse sequence and following $\phi_2$ corresponds to the CID pathway.  In this case, adiabaticity is achieved if $\mathcal{A}_{\rm{CTAP}}(t)\ll1$ for all $t$.  This method has the advantage that an estimate for the minimum pulse time required is obtained without explicitly integrating the time dependence.  The disadvantage is that it provides no information on the fidelity (probability of successful transfer) of a given pulse sequence, which still requires numerical integration.  Fig.~\ref{fig:adiab} shows $t_{\rm{max}} \mathcal{A}_{\rm{CTAP}}(t)$ as a fraction of the pulse time for each of the viable pulse sequences. Looking at the maximum of the adiabaticity, we expect that pulse \texttt{IV} can be applied up to 4 times faster than pulse \texttt{I}.  It is interesting to note that the adiabaticity predicts a difference in the behaviour of pulses \texttt{I} and \texttt{II}, despite the similarity of the corresponding eigenspectra.  We also see that the maxima of $\mathcal{A}_{\rm{CTAP}}(t)$ are not at the middle of the pulse sequence, in contrast to the FS model. 

\section{Central well population}\label{sec:well2}
One of the defining features of CTAP protocol in the FS formalism is that the central well is unpopulated in the adiabatic limit.  One key question is whether this feature is preserved when including the effects of realistic 1D potentials.  Using the finite state formalism, the population is predicted to be exactly zero for infinitely slow evolution due to the precise cancellation of the components of the wavefunction which correspond to the middle well.  Therefore the fraction of population in the central well due to the finite wells represents an important difference between the FS and SW descriptions.  When performing simulations on the effects of decoherence and fabrication defects\cite{Kamleitner:08,Ivanov:04}, it is essential to include these deviations from the FS picture as these represent one of the principle error modes.  Non-adiabatic corrections will also introduce some population in higher lying states (as seen in the conventional Landau-Zener problem) although this effect is replicated, at least qualitatively, in the FS formalism.  

\begin{figure} [tb!]
\centering{\includegraphics[width=9cm]{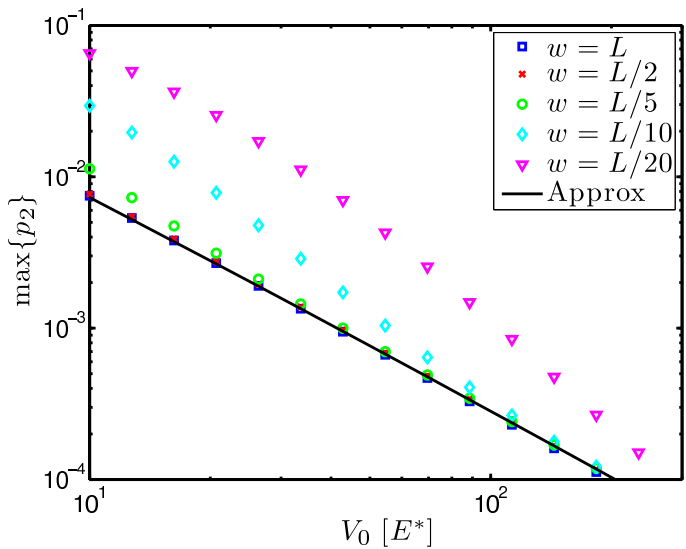}} 
\caption{Occupation probability of the central region as a function of barrier height and barrier width.  Note that for sufficiently high barriers, the occupancy is largely independent of barrier width.  Even with the barrier at its minimum value during a CTAP pulse, the population of the central well and the barriers is less than 0.1 for a barrier width of $w=L/20$.\label{fig:2nd_well_occupancy}}
\end{figure}

Using the exact diagonalisation over a finite lattice, we can see that for a finite barrier the population is not zero, even in the perfectly adiabatic limit.  This non-zero population results in the system now being susceptible to decoherence and other effects which depend on the population of well 2.  As the CID pulse should transfer an electron from well 1 to 3 without populating well 2 (according to the FS model), we can define the probability of deviation from the ideal result as the maximum occupancy of the central well and surrounding barriers
\begin{equation}
p_{2}=1-p_{\rm{well1}}-p_{\rm{well3}}=\int_{-L/2-w}^{L/2+w} \psi^*\psi dx.
\end{equation}
In Fig.~\ref{fig:2nd_well_occupancy} we plot this probability as a function of both barrier height $V_1=V_2=V_0$ and barrier width $w$.  This plot shows that the population is exponentially suppressed but non-zero for finite barrier, as foreshadowed earlier.

We can also solve this problem analytically in the limit of large barriers.  In the limit that the barriers are high or wide (or both), the major contribution to $p_{2}$ comes from the exponential decay of the wavefunction as it enters the barriers (rather than population in the central well itself).  We can therefore model this using the solution for a particle in a box with one infinite wall and one finite wall.  In this case, the solution is a transcendental equation,
\begin{equation}
\frac{k_1}{k_2} = -\tan(k_1 L)
\end{equation}
where $k_1=\sqrt{2mE/\hbar^2}$, $k_2=\sqrt{2m(V_0-E)/\hbar^2}$ and $L$ is the width of the well.  The central region probability is then approximately the fraction of the particle distribution function which is inside the barrier.  This approximate solution is also plotted in Fig.~\ref{fig:2nd_well_occupancy} and is valid for $w \times V_0 \gtrsim 10 LE^*$, i.e.\ barrier areas greater than about $10$ in normalised units. 

\section{Time evolution}\label{sec:timeevol}
To calculate the time evolution of an electron in a spatially and temporally varying potential, we use the Crank-Nicolson method\cite{Crank:47,Crank:96,Wong:92} to approximate the propagator over a time period $\Delta t$.  The time evolution of the system is then expressed as
\begin{equation}
\Psi(x,t+\Delta t) = U(\Delta t) \Psi(x,t),
\end{equation}
where the propagator in Cayley form\cite{Goldberg:67,Wong:92} is
\begin{equation}
U(\Delta t) \approx \frac{2-i(\Delta t)\hat{H}}{2+i(\Delta t)\hat{H}}.
\end{equation}
This allows the time evolution of the system to be calculated for an arbitrary combination of initial state, potential and pulse scheme.  While this method is much more computationally intensive than solving the same problem using the FS formalism, it does describe the effects of finite barriers, spatially distributed wavefunctions and arbitrary potential profiles. 

\begin{figure} [tb!]
\centering{\includegraphics[width=9cm]{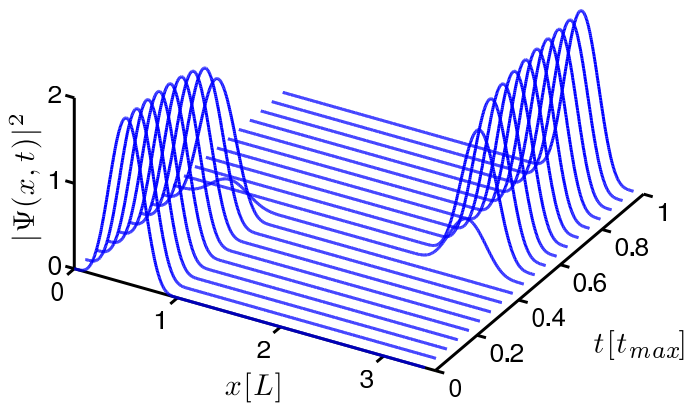}} 
\caption{Example trace showing evolution of the probability distribution as a function of time during a CID pulse for pulse \texttt{II}.  Qualitatively similar behaviour is observed for the other pulse sequences.\label{fig:psitexample}}
\end{figure}

In Fig.~\ref{fig:psitexample} we plot a series of snapshots showing the probability distribution, $|\Psi(x,t)|^2$, as a function of time during the application of pulse sequence \texttt{II}.  As the barriers are lowered, the particle starts to penetrate into the barriers (indicated by a spreading of the wavefunction) and then ultimately move from well 1 to well 3 with only an exponentially suppressed population in the well 2.

\begin{figure} [tb!]
\centering{\includegraphics[width=9cm]{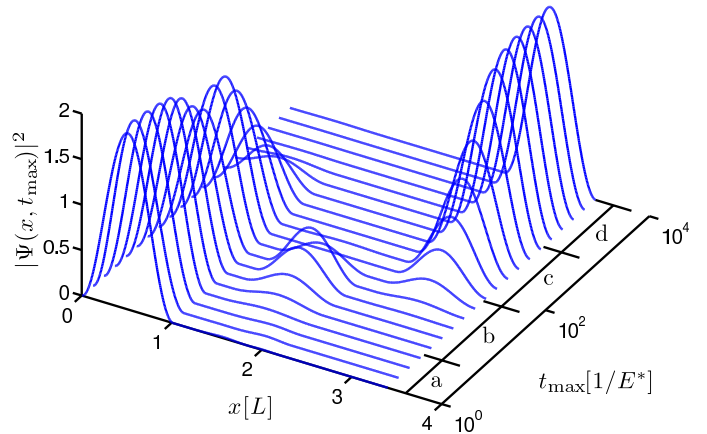}} 
\caption{Plot showing the final state of the system as a function of the total time of pulse \texttt{II}.  Note the four key regions (a), (b), (c) and (d) described in the text, which correspond to four regions of qualitatively different behaviour.\label{fig:tmaxwfnc}}
\end{figure}

To study the approach to the adiabatic limit, in Fig.~\ref{fig:tmaxwfnc} we plot the final population distribution $|\Psi(x,t_{\rm{max}})|^2$ as a function of $t_{\rm{max}}$, using pulse \texttt{II} scaled appropriately for each $t_{\rm{max}}$.  We identify four regions based on the systemÕs qualitative properties, the simplest being regions (a) and (d).  In region (a), the pulse sequence is applied so quickly that the system doesn't have time to respond, resulting in no evolution.  In region (d), the system evolution is slow enough that the entire transfer is adiabatic, resulting in complete transfer from well 1 to 3.  In the intermediate regions (b) and (c), the transfer is slow enough to allow evolution but still non-adiabtic in that the other eigenstates are populated during the pulse, resulting in imperfect transfer.  It is interesting to note that in region (b), all 3 wells are populated at the end of the transfer, whereas in region (c) the final state is only composed of population in well 1 and 3.  This population asymmetry arises because the spacing between the eigenstates is non-uniform, due to the offset in the central well potential.  Referring back to Fig.~\ref{fig:eigenspec1}, we see that the energy gap between the 2nd and 3rd eigenstates is larger than that between the 1st and 2nd.  We can therefore interpret region (b) as the point at which the transfer is fast enough to populate all three eigenstates whereas the slower pulse in region (c) can only populate the lower two states.  As the lower two states are only adiabatically connected to the 1st and 3rd wells, this results in the observed population distribution. 

\begin{figure} [tb!]
\centering{\includegraphics[width=9cm]{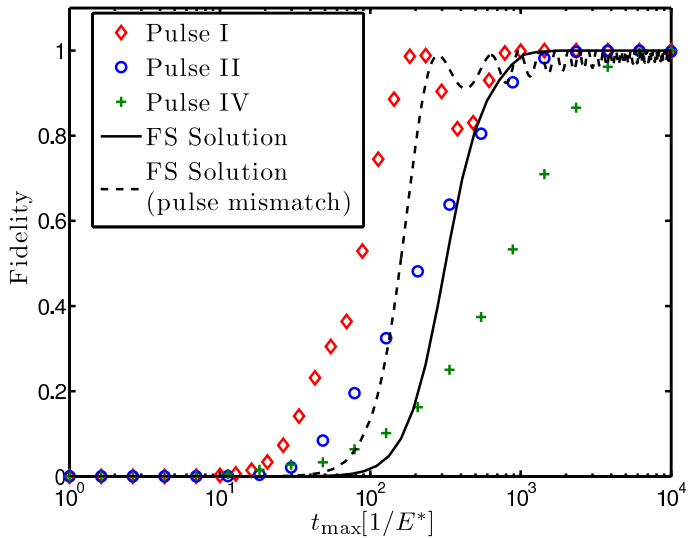}} 
\caption{Fidelity of the pulse as a function of the total pulse time, shown for three different pulses.  The FS solution to this problem is also shown (solid line) with Gaussian overlapping pulses as well as the same problem with a timing mismatch (dashed line) such that the time separation between the pulses is doubled.  The oscillations in the fidelity stem from this imperfect overlap in the pulses.\label{fig:fidvstmax}}
\end{figure}

To compare the various pulsing schemes, we plot the probability of successful transfer as a function of the pulse time ($t_{\rm{max}}$) in Fig.~\ref{fig:fidvstmax}, following the similar analysis in Ref.~\onlinecite{Greentree:04}.  The ordering is as expected from Fig.~\ref{fig:adiab} with a pulse time of $t_{\rm{max}}\gtrsim 10^3$ allowing adiabatic transfer for pulse \texttt{II} whereas pulse \texttt{IV} requires approximately 4 times longer.  While the fidelity for pulse \texttt{II} \& \texttt{IV} are monotonically increasing functions, the fidelity function for pulse \texttt{I} has a noticeable oscillation.  This is consistent with pulse \texttt{I} being an approximation to pulse \texttt{II}.  Previous work has shown that the fidelity function of a CTAP pulse (using Gaussian modulation) is strongly dependent on the spacing between the pulses\cite{Greentree:05}.  In this case the approximation used to derive pulse \texttt{I} is only strictly valid near the minimum barrier height (maximum tunnelling rate) whereas the overlap between the pulses is strongly controlled by the value of the barriers at $t=t_{\rm{max}}/2$.  This variation in the overlap leads to oscillations in the fidelity function.

For comparison, the solution of the same problem using the FS formalism with appropriate energies is also shown in Fig.~\ref{fig:fidvstmax}.  For the pulsing scheme given by Eq.~(\ref{eq:gaussianpulses}), the fidelity increases monotonically (solid line),  although the maximum gradient of this line is larger than that predicted using the SW formalism.  The fidelity obtained using the FS formalism is also plotted for a Gaussian barrier modulation where the separation between pulses has been doubled (dashed line), illustrating the onset of oscillations in the fidelity due to mismatch in the pulse timing.   

The choice of pulse modulation is therefore important with the maximum transfer speed ultimately controlled by the adiabatic criteria for the transfer.  To optimise the transfer rate, it would therefore be useful to design a pulse sequence which minimises the adiabatic criteria while still connecting the initial and final states via an adiabatic pathway~\cite{Malinovsky:97}.  In practice, this optimisation would also need to take into account any restrictions on the rate at which the barriers can be varied as well as the maximum and minimum values the barrier height can take.

\section{Conclusion}
Recent progress in theoretical and experiment control of quantum systems has highlighted the link between quantum optics and coherent solid-state devices.  Many of the key developments have benefited from the close mathematical and conceptual relationship between these fields.  This has allowed a fast uptake of ideas, where a concept in one field can be quickly mapped to the other.  We have investigated aspects of this link, specifically where they apply to the use of adiabatic passage techniques in solid-state systems.  The characteristics of the eigenspectrum and evolution of an electron in a triple-well system was studied for different time varying potentials.  These results are directly applicable to electron shuttling in confined electron systems such as quantum dots or between donor sites.

While the analogy between the two formalisms studied is strong enough to guide future work, care must be taken, especially when using realistic potential profiles to predict the outcomes of future experiments.  For more sophisticated and accurate calculations, it will be ultimately necessary to include more physics than is contained in the finite state approximation.  This is particularly evident in the example considered in this paper when trying to determine the probability of success for various pulse designs and switching times at the limit of the adiabatic window.  At this point the spatial wave model predicts both qualitatively and quantitatively different behaviour to the finite state model, in terms of both transfer fidelity and failure modes.  In these situations, the solution of the spatially varying wavefunction is essential for an accurate description of the system dynamics.

The goal of our work is to assess the validity of the usual (quantum optics motivated) approach to the CTAP process where the spatial dependence of the phenomenon is ignored.  Our model analysis, using the explicit solution of the time-dependent Schr\"odinger equation, demonstrates the validity of the usual CTAP approach, pointing out at the same time some of the limitations arising from the neglect of the spatial-dependence in the standard (and extensively used) CTAP theory.  We are particularly interested in the application of CTAP in the solid state quantum information processing in quantum dot or semiconductor donor systems.  Since such solid state CTAP experiments are still very far in the future, we have used a minimal model so as not to complicate the conceptual issues by unnecessary numerical
details.  A real experimental situation will necessitate going beyond this minimal model, which may turn out to be complicated in real solid state systems.

\section*{Acknowledgments}
The authors wish to acknowledge useful discussions with A.~M.~Martin and D.~P.~George.  JHC acknowledges the support of the Alexander von Humboldt Foundation and SDS acknowledges research
support from the LPS-NSA.  ADG is the recipient of an Australian Research Council Queen Elizabeth II Fellowship (DP0880466) and LCLH is the recipient of an Australian Research Council Australian Professorial Fellowship (DP0770715).  This work was supported in part by the Australian Research Council, the Australian Government, the US National Security Agency and the US Army Research Office under contract number W911NF-04-1-0290.   

\bibliography{ctap_potentials}

\end{document}